\documentclass[11pt,a4paper]{article}
\pdfoutput=1
\usepackage[utf8]{inputenc}
\usepackage{jheppub}  
\usepackage{amsfonts,amsbsy,amsmath,amssymb}

\newcommand{\Tr}{\text{Tr}}

\newcommand{\be}{\begin{equation}}
\newcommand{\ee}{\end{equation}}
\newcommand{\ba}{\begin{array}}
\newcommand{\ea}{\end{array}}
\newcommand{\baa}{\begin{array}}
\newcommand{\eaa}{\end{array}}
\newcommand{\bea}{\begin{eqnarray}}
\newcommand{\eea}{\end{eqnarray}}
\newcommand{\half}{\frac{1}{2}}

\newcommand{\W}{\mathcal{W}}
\newcommand{\V}{\mathcal{V}}
\newcommand{\Op}{\mathbf{O}}
\newcommand{\bW}{\bar{\mathcal{W}}}
\newcommand{\Ga}{\Gamma^{(a)}}
\newcommand{\Gda}{\Gamma^{(a) \dagger}}

\newcommand{\II}{\mathbb{I}}

\newcommand{\del}{\boldsymbol{\delta}}
\newcommand{\WZERO}{Y}
\newcommand{\hG}{\boldsymbol{\hat \Gamma}}
\newcommand{\hGSa}{\hG^{(a)}}
\newcommand{\hGSo}{\hG^{(1)}}
\newcommand{\hGSt}{\hG^{(2)}}
\newcommand{\f}{\bar{f}}

\title{Constructing SU(N) fractional instantons}

\author[a,b]{Antonio Gonz\'alez-Arroyo}

\affiliation[a]{Instituto de F\'{i}sica Te\'orica UAM-CSIC, Nicol\'as
  Cabrera 13-15, Universidad Aut\'onoma de Madrid, E-28049 Madrid, Spain}
\affiliation[b]{Departamento de F\'{i}sica Te\'orica, 
M\'odulo 15,  Universidad Aut\'onoma de Madrid, Cantoblanco, E-28049 Madrid, Spain}

\emailAdd{antonio.gonzalez-arroyo@uam.es}

\abstract{ We study self-dual SU(N) gauge field configurations on the 4 torus with
twisted boundary conditions, known as fractional instantons. Focusing
on the minimum non-zero action case, we 
generalize the constant field strength  solutions discovered by `t
Hooft and valid for certain geometries. For the general case, we 
construct the vector potential and field strength in a power series
expansion in a deformation parameter of the metric. The next to
leading term is explicitly computed. The methodology is an
extension of that used by the author for SU(2) fractional instantons
and for vortices in two-dimensional Abelian Higgs models. Obviously,
these solutions can also be seen as self-dual configurations in
$\mathbb{R}^4$ having a crystal structure, where each node of the
crystal carries a topological charge of $1/N$. 
}


\preprint{%
{\flushright
IFT-UAM/CSIC-19-138\\
FTUAM-19-19\\
}}

\begin{document}

%\PACS{ 11.15.Pg,   11.15.Ha,  11.10.Nx }

\maketitle

\section{Introduction}
The study of solutions of the classical equations of motions from the
perspective of quantum field theory studies has a long history (see
here~\cite{Coleman:1985rnk, Rajaraman:1982is} some books written on the subject). For the case of 
gauge theories a pioneering role has been played by the BPST instanton
solution~\cite{Belavin:1975fg} of the Yang-Mills euclidean equations of motion.
Its interpretation and relevance  in the quantum field theory setting 
as dominating  tunneling trajectories was
clarified by Polyakov~\cite{Polyakov:1975rs}. As in all solutions to partial
differential equations, boundary conditions matter. In the case of
instantons the condition of finite action is equivalent to the
compactification of $\mathbb{R}^4$ into $S_4$. This automatically
brings in the study of the topology of the bundles on compact
manifolds of great interest to
mathematicians~\cite{donaldson1990geometry}. These
bundles can be classified according to Chern classes. In particular,
the stability of the instanton solution is a consequence of it
possessing a non-trivial second Chern-number, also known as instanton
number. A similar phenomenon happens in two-dimensional abelian gauge 
theories  with the vortex solution~\cite{abrikosov, Nielsen-Olesen} and
its connection with the first Chern number.  The history of these
topics is beautiful and very rich but unfortunately we cannot review
it here. This paragraph serves to put our work in context but we will
now focus on aspects more directly related to our work. 

The study of gauge fields on the four-dimensional torus within the Physics
literature was initiated by `t Hooft. The torus has many advantages as
a compact manifold since it is compatible with a flat metric and
allows to respect a group of translations of practical interest.
Furthermore, numerical studies of gauge theories are almost always
performed  on the torus. What `t Hooft put forward is a new class of
topological sectors of SU(N) gauge  bundles on the torus, which he called
twisted boundary conditions~\cite{hooft_em,hooft_emp,hooft_q}. These sectors are characterized by an
antisymmetric tensor of integers modulo $N$. These integers
can be interpreted as abelian fluxes through the faces of the torus. They are a
remnant of the first Chern numbers of a U(N) gauge theory when
projecting onto SU(N), or as mathematicians see it, as obstructions to
lifting an SU(N)/Z\_N bundle to a SU(N) one. These questions were clarified
by van Baal in his thesis and early papers~\cite{vanBaal:1984ra,vanbaal2}. One of the
interesting aspects discovered by `t Hooft was the connection of the
twist fluxes with the instanton number. It turns out that for certain
twist tensors the instanton  number is no longer an integer. All this
is perfectly understood in mathematical terms as a result of the work
of van Baal and others~\cite{Sedlacek:1982cd, Nash:1982kp}. An
explicit construction of the bundles implementing all possible values
of the instanton number and twist tensor can be seen in my lectures
given several years ago~\cite{peniscola}.

The search of new classes of Yang-Mills classical solutions  on the torus 
having fractional 
topological charge was initiated by `t Hooft~\cite{thooft}. We
will refer to all self-dual solutions as {\em fractional instantons},
although the reader is warned about the use of different names  in
the literature for the same objects. `t Hooft analytic solutions are
constant field strength solutions which become self-dual (and hence
stable) only when the ratio of certain areas of the faces of the torus
become certain rational numbers. The solutions have a  somewhat abelian 
character since the electric and magnetic fields are aligned along a
single direction in Lie Algebra space (and hence commute among
themselves). However, using  numerical methods one can obtain
fractional instantons for a wide range of torus sizes including the
Hamiltonian $T_3\times \mathbb{R}$ geometry~\cite{GarciaPerez:1989gt,
GarciaPerez:1992fj}. These instantons are lumpy structures having a center in
space and time and non-commuting non.constant electric and magnetic
fields. The numerical technique reflects the methods used in
mathematical approaches to the subject: gradient flows. Starting from
non self-dual configurations on the twisted bundle, the flow converges in some
cases to self-dual non-singular configurations. 

In commenting about the relevance of these fractional instantons to
the dynamics of Yang-Mills fields we should clarify a few points that
unfortunately some fraction of the scientists seem to ignore. Boundary
conditions are necessary only to stabilize the solutions. The
classical equations of motion are {\em local} equations satisfied at
each space-time point and this makes these structures relevant even if
we modify the boundary conditions. Let us put two very simple examples
to explain this point. Consider first the one-dimensional scalar field
theory with a double well potential. If we put antiperiodic boundary
conditions in time, there is a stable classical solution known as the
kink. If we change the boundary conditions to periodic, the
kink is no longer a stable solution, but configurations with kinks and 
antikinks provide unstable classical configurations of great dynamical
relevance. Another quite different example is that of $Q=1$ instanton
on the torus without twist. This configuration is
unstable~\cite{Braam:1988qk}. The
configuration will shrink in size with gradient flow tending towards a 
singular solution. Does this mean that compactifying space-time on a
torus will make instantons absent in the Yang-Mills  vacuum? Obviously
not. Quantum fluctuations will produce instantons since boundary
conditions only affect the total action through a term proportional to
the boundary, which is subdominant with respect to the action or
entropy of the bulk. Coming back to fractional instantons, twist only
plays an stabilizing role. Since twist is a kind of flux  modulo N,  if we
glue together  several copies of the torus in various directions we
produce a larger torus with no twist. The resulting configuration is
still a self-dual solution (hence minimum action) but the topological 
charge (which is not modulo N) is  larger and hence the moduli space
grows accordingly. This means that there are deformations that cost no
action that destroy the periodicity under the small period. It is
obvious that for small enough deformations the configurations will still look like a
collection of lumps carrying fractional topological charge. Curiously,
the moduli space of all self-dual solutions has precisely the same
dimensionality  as a four-dimensional gas of fractional instantons of
minimum action $(8\pi^2/N)$. This is the same as in abelian gauge
theories in two dimensions, in which the moduli space is precisely
given by the  configuration space of a two-dimensional gas of
vortices~\cite{taubes2}.

Do these fractional instantons play a dynamical role in the Yang-Mills
vacuum? Instantons are important in solving the U(1)
problem~\cite{tHooft:1976rip} and producing a non-zero topological charge
density, but cannot explain other phenomena such as confinement. Very
early on some researchers proposed that instantons can dissociate into
some constituents in a dense scenario which could be responsible for
Confinement~\cite{Callan:1977qs}. Unfortunately at the time the
only possible candidates were some singular configurations known as
merons~\cite{deAlfaro:1976qet}. Our proposed confinement
scenario~\cite{GarciaPerez:1993jw, GonzalezArroyo:1995zy} claims that the idea is basically
correct but replacing the old singular meron solutions by the regular,
smooth, self-dual fractional instantons. The idea arose quite
naturally when pursuing a program initiated by
Luscher~\cite{Luscher:1982uv}
of trying to use the spatial volume as an interpolating parameter
between the perturbative femtoworld and the large volume confinement
regime. Fractional instantons appear naturally as non-perturbative
weak coupling solutions whose effect is to approach the system towards 
the confinement regime~\cite{GarciaPerez:1993ab}. Further studies done by our
group in Madrid hinted on the presence of fractional instantons on the
large volume lattice configurations~\cite{GonzalezArroyo:1995ex} and
showed that an
artificially created gas of fractional instantons leads to Wilson
loops satisfying the area law~\cite{GonzalezArroyo:1996jp}. At about the same
time Zhitnitsky~\cite{Zhitnitsky:1989ds, Zhitnitsky:1991eg} advocated the existence and relevance of fractional
topological charge objects to explain the N dependence of the free
energy in the presence of a $\theta$ parameter.

Coming back to fractional instantons at the classical level, one
important difficulty is that there are no analytic formulas for the
vector potentials or the field strengths beyond the special solutions
found by `t Hooft. Of course, the same happens for the vortex and
multivortex solutions in abelian two dimensional field theory. Thus,
it became as a wonderful surprise when analytic formulas 
were obtained for $Q=1$ non-trivial holonomy calorons in which the
dissociation mechanism is explicit~\cite{Lee:1998vu, Lee:1998bb,Kraan:1998kp,Kraan:1998pm}. The
moduli space of these solutions nicely interpolates smoothly between
a single ordinary instanton lump and a set of $N$ local lumpy structures carrying fractional
topological charge. These caloron constituents  are intimately connected to
fractional instantons. Indeed, for large separations these $S_1\times\mathbb{R}^3$ 
fractional topological charge caloron components 
 can be seen as a one-dimensional periodic array of minimum action
fractional instantons. These fractional instanton constituents can
also be arranged into two dimensional doubly periodic
sheets~\cite{GonzalezArroyo:1998ez, Montero:1999by, Montero:2000pb} 
that make up solutions in $T_2\times \mathbb{R}^2$, as in the
three-dimensional $T_3\times \mathbb{R}$ solutions mentioned earlier.
All this makes a unified picture of these arrays of fractional
instantons~\cite{GarciaPerez:1999bc, GarciaPerez:1999hs}, connected to each other by Nahm
transformations~\cite{Nahm:1979yw, Corrigan:1983sv, Braam:1988qk,
Schenk:1986xe, GonzalezArroyo:1998ia}. 

In search for analytical expressions of other types of general fractional
instantons we developed a strategy based on deforming away from `t
Hooft constant field strength solutions~\cite{su2paper}. A
perturbative expansion on the deformation parameter  arises naturally
and we were able to show that the equations can be solved order by
order in a sequential fashion and to compute  the leading orders in
the expansion. Our analytical formulas reproduced the numerical
solutions obtained for small deformations, confirming the validity of
our results. Going beyond the first few orders is necessary if one
wants to obtain a good approximation to general torii. Investigating
this matter we realized that the deformation idea is very general.
Indeed, something very similar happens for two-dimensional abelian
gauge theories on the torus. A constant  field strength solution
exists for a particular value of the area. Deforming the area one
obtains a hierarchy of equations that can be solved order by order to 
provide analytical formulas for the critical vortices on the
torus~\cite{GonzalezArroyo:2004xu}.
Indeed, in this simpler case we were able to compute up to order 51 in
the deformation parameter, which allows to reproduce nicely the
critical vortex in $\mathbb{R}^2$ (infinite area). Even more so, the
procedure allows the computation of multivortex solutions at all
points in the moduli space. This leads to an analytic control of
features such as vortex-vortex scattering~\cite{GonzalezArroyo:2006gn} or
quantum corrections to the multivortex energies~\cite{Ferreiros:2014mca}. The
existence of a critical area is completely general in these
two-dimensional  abelian-Higgs systems as discovered in the thesis of
mathematician Steven Bradlow~\cite{Bradlow:1990ir}. Thus we named the
expansion in the deformation parameter   as Bradlow parameter expansion.

Many of the previous works, including our deformation perturbative
approach~\cite{su2paper} focused on SU(2). This was mostly driven
by simplicity and/or  computational resources. However, some numerical
work was done for SU(N) fractional
instantons~\cite{GarciaPerez:1997fq,Montero:2000mv, Montero:2000pb}
showing that the ideas and results extend to all values of $N$.
Indeed, large N was always in the origin of our interest in twisted
boundary conditions. Fractional instantons have free energies that
survive the large N limit. Twist also plays a major
role~\cite{TEK1,TEK2} in preserving enough center symmetry at weak
coupling, a necessary ingredient for the validity of volume reduction
at large N~\cite{EguchiKawai}. Thus,   we felt the necessity of
extending  our previous analytic expansion to SU(N). This is indeed the 
main purpose of this paper. 

The motivation for the extension has emerged from a recent interest in
fractional instanton solutions emerging from a different perspective.
This comes in the spirit of the ideas of resummation of the
perturbative expansion and the proposed idea of
Resurgence(See Ref.~\cite{Dunne:2016nmc} for a recent review). The claim is that even in the case of
non-Borel summable expansions one can use the perturbative expansion
to reconstruct all non-perturbative phenomena. A huge literature has
emerged which we cannot review here. The main connection with our
program is that in certain simpler systems fractional soliton solutions 
of various kinds have been found to be relevant in interpreting the
singularity structure in the Borel plane~\cite{Cherman:2013yfa}. The natural
candidate to extend this phenomenon to four-dimensional gauge theories
are precisely the fractional instantons. Hence, we found the courage
to extend the construction to SU(N), hoping our formulas will be of
some help to other researchers. In so doing we have also generalized
some of the steps that were previously carried only for special cases.

The lay-out of the paper is as follows. In the next section we collect
several results about fractional instantons. Some of the concepts
mentioned in this introduction will be put in mathematical terms. The
following section is devoted to constant field strength solution. We
actually generalize the construction done by `t Hooft both in the
group structure as in the geometrical setting. We will focus only on
minimal action instantons given their unique character. The following section
explains the philosophy of the deformation technique and show that it
leads to hierarchy of equations.  Section~\ref{orderepsilon} is devoted to the
computation of fractional configuration to  first order in the  deformation 
parameter. In section~\ref{higherorders} we show the basic ingredients to
extend the calculation to higher orders. Finally, in the last section
we present our conclusions and explain how our results can be extended
and/or used to compute other interesting quantities such as fermion
zero modes.

\section{Fractional Instantons}
\label{general}
\newcommand{\T}{\mathcal{T}}
In this section we recall some general facts about gauge fields on a
four-dimensional torus. The torus $\T$ is given as the quotient space
$\mathbb{R}^4/\Lambda$ where $\Lambda$ is a discrete group of
translations generated by 4 linearly independent vectors $e_\alpha$ for $\alpha \in
\{0,1,2,3\}$. In our favourite presentation we introduce SU(N) gauge fields
as connections in an SU(N) vector bundle. The bundle itself is defined
by its transition functions, which is given by a homomorphism from
$\Lambda$ to the space of gauge transformations. In this way we
guarantee that all gauge-invariant quantities are well-defined on the
torus. In an specific trivialization all we need to do is to associate
an SU(N) matrix to each generator 
\be
e_\alpha \longrightarrow \Omega_\alpha(x)
\ee
so that each section of the bundle $\Psi(x)$ transforms as
\be
\Psi(x+e_\alpha)=\Omega_\alpha(x) \Psi(x)
\ee
Consistency then demands that 
\be
\Omega_\alpha(x+e_\beta)\Omega_\beta(x)=
\Omega_\beta(x+e_\alpha)\Omega_\alpha(x)
\ee

The space of bundles can be classified into topologically inequivalent
sectors by means of the Chern classes. The first Chern class
integrated over non-trivial 2-cycles gives the first Chern numbers.
Thus, to  each face of the torus we can associate a number which can
be interpreted as the flux through that face. However, for SU(N),
these numbers are all zero. We then have the second Chern class, which
when integrated over the full space gives the second Chern number, 
 instanton number or topological charge $Q$. This number is known to be an integer. The
best way to compute this number is by introducing a connection
$A=A_\mu(x) dx^\mu$ on the bundle with its corresponding curvature
2-form $F$. The instanton number is given by 
\be
 Q= \frac{1}{8 \pi^2} \int_\T \mathrm{Tr}(F \wedge F)
\ee
Notice, however, that the number is a property of the bundle encoded in its
transition matrices. 

`t Hooft realized that in pure gluodynamics the consistency conditions 
can be relaxed to the form~\cite{hooft_em,hooft_emp} 
\be
\Omega_\alpha(x+e_\beta)\Omega_\beta(x)= z_{\alpha \beta}
\Omega_\beta(x+e_\alpha)\Omega_\alpha(x)
\ee
where $z_{\alpha \beta}=\exp\{2 \pi i n_{\alpha \beta}/N\}$ and
$n_{\alpha \beta}$ is an antisymmetric tensor of integers modulo $N$.
The connection is still  well-defined because it is insensitive to a
transformation by an element of the center $Z_N$. These modified consistency
conditions were called twisted boundary conditions by `t Hooft. The
6 independent integers of the twist tensor $n_{\alpha \beta}$,
can be written as 2 integer 3-vectors $\vec{k}$ and $\vec{m}$
($k_i=n_{0 i}$ and $m_i=\epsilon_{i j k} n_{j k}/2 $) defined modulo $N$. Their integer
character  shows that they characterize topologically inequivalent bundles. 
`t Hooft also realized that the instanton number is related to these
vectors as follows
\be
Q=-\frac{\vec{k}\vec{m}}{N}+ \mathbb{Z}
\ee
We then see that for non-orthogonal twists ($\vec{k}\vec{m}\ne 0 \bmod
N$) the instanton number becomes fractional. This apparent puzzle was
clarified by Pierre van Baal in his thesis~\cite{vanBaal:1984ra,vanbaal2}. In reality
we are  constructing an $SU(N)/Z_N$ bundle, and one should write the
transition matrices in a center-blind representation as the adjoint.
Twist becomes then an obstruction to lifting the bundle to one in SU(N). Another way to
look at twist is by starting with a U(N) bundle and projecting it down
to SU(N)~\cite{Lebedev:1988wd}. It is then clear how the first Chern number of the original 
bundle relates to the twist of the SU(N). This clarifies the
interpretation of twist as flux modulo $N$. For a  more extensive
description of the preceding, the reader can also consult the author's
lectures~\cite{peniscola}, which includes an explicit construction of
twist matrices for all values of the twist and instanton number for $N>2$.  

To generate the  dynamics of gauge fields one introduces the
Yang-Mills action functional
\be
 S= \frac{1}{2 g^2} \int_\T dx \ \Tr( F_{\mu \nu} F^{\mu \nu})
\ee
Implicitly this demands the introduction of a metric on the torus
($dx$ stands for the corresponding volume form), although one
frequently takes it to be the euclidean metric.  Here, we will stick
to this case. However, as we will see, the expression of the constant
metric tensor in a given coordinate system will play a fundamental
role in what follows.  

The action is bounded
from below by a multiple of the absolute value of the topological
charge~\cite{bogomolny}
\be
S\ge \frac{8 \pi^2 }{g^2} |Q|
\ee
This Bogomolny bound is saturated by self-dual or anti-self-dual
configurations. These configurations are called instantons (for the
self-dual $Q>0$ case), whose
first representative is the  celebrated BPST
instanton~\cite{Belavin:1975fg}
having $Q=1$ on $S_4$ or $\mathbb{R}^4$. For the torus case the
possible solutions for non-integer $Q$ are called fractional
instantons. Their existence has been established mathematically in
some cases~\cite{Sedlacek:1982cd}. One can start by a configuration in 
each sector (which is known to exist) and then apply a gradient 
flow to minimise the action. The limit does not necessarily exist (as
happens for $Q=1$ and $\vec{k}=\vec{m}=0$) because the limiting
configuration can be singular. The method has been used successfully to
obtain numerically precise approximations to the fractional instanton
configurations for certain
geometries~\cite{GarciaPerez:1989gt,GarciaPerez:1992fj,Montero:2000mv}.   

The fractional instanton configurations are not unique, but depend on
$4|Q|N$ real parameters as dictated by the index theorem. Particular
interest is attributed to the lowest action fractional instanton
having topological charge $|Q|=1/N$. Apart from acting as building
block for the higher topological charge solutions, it is essentially
unique, since its $4$ moduli parameters are associated to space-time
translations. Notice that in this case the action remains finite in
the large $N$ limit and given by $8 \pi^2 / (g^2 N)$. 

Before describing the analytic construction of these solutions we
should mention that the torus and the twisted boundary conditions are
only auxiliary tools in their identification. The configurations can
be seen as configurations in $\mathbb{R}^4$ satisfying certain
periodicity conditions. They are still solutions of the classical equations of
motion (with euclidean signature) although with infinite action
(finite action over each cell). It is also important to realize that
since twist fluxes are additive modulo $N$, fractional instantons also
give rise to classical solutions on the torus  with vanishing
twist $\vec{k}=\vec{m}=0$ and integer topological charge. These
configurations look  very different to a collection of $Q=1$
instantons. 

In the next section we will present all constant field strength
fractional instanton solutions, which are valid  for specific torus
sizes, thus  generalizing `t Hooft construction~\cite{thooft}.

\section{Constant field strength fractional instantons}
\label{constant}
`t Hooft succeeded in obtaining analytical solutions for some
fractional instantons~\cite{thooft}. A good deal of importance comes
from choosing the transition matrices. He used a hybrid between the
abelian and the twist-eating  matrices~\cite{Groeneveld:1980tt,
Ambjorn:1980sm}:
\be
\Omega_\alpha(x)=e^{i \pi \hat\omega(e_\alpha,x) T } \begin{pmatrix} 
\Gamma^{(1)}_\alpha & 0 \cr 0 & \Gamma^{(2)}_\alpha \end{pmatrix} 
\ee
where $\Gamma^{(a)}_\alpha$ are constant SU($N_a$) matrices
satisfying 
\be
\label{twisteating}
\Gamma^{(a)}_\alpha \Gamma^{(a)}_\beta= e^{2 \pi i n^{(a)}_{\alpha
\beta}/N_a}  \Gamma^{(a)}_\beta \Gamma^{(a)}_\alpha 
\ee
and $T$ is a hermitian traceless matrix commuting with all the
$\Omega_\alpha(x)$. Explicitly we have 
\be
T=  \begin{pmatrix} \frac{\II_1}{N_1} & 0 \cr 0 & -\frac{\II_2}{N_2} \end{pmatrix}
\ee
with $\II_a$ the $N_a\times N_a$ identity matrix. We have split the
space into two blocks such that $N_1+N_2=N$.  Finally,
$\hat\omega(x,y)$ is a an antisymmetric bilinear form. Imposing the twisted boundary conditions one
concludes that 
\be
n_{\mu \nu}=n^{(1)}_{\mu \nu}+n^{(2)}_{\mu \nu} \quad \Leftrightarrow
\quad \vec{k}=\vec{k}^{(1)}+ \vec{k}^{(2)} \quad  ; \quad
\vec{m}=\vec{m}^{(1)}+ \vec{m}^{(2)}
\ee
and 
\be
\label{defDelta}
\hat\omega(e_\alpha,e_\beta)= \frac{\Delta_{\alpha \beta}}{N}\equiv
\frac{n^{(2)}_{\alpha \beta} N_1 - n^{(1)}_{\alpha \beta} N_2}{N}
\ee
We have used the freedom to redefine $n_{\alpha \beta}^{(a)}$ modulo
$N_a$, to write these equations as exact and not modulo integers.  
We recall (see ~\cite{peniscola}  and references therein) that the existence of  
solutions to Eqs.~\eqref{twisteating} implies 
\be
\label{conds}
\vec{k}^{(a)}\vec{m}^{(a)} =0 \bmod N_a 
\ee

Associated to the aforementioned twisted transition matrices there is
a natural constant field strength connection. The vector potential
one-form is given by $\hat A=\pi \hat\omega(x,dx) T $ with field
strength $\hat F=2
\pi \hat\omega(dx,dx) T$. We can use this connection to compute the
topological charge 
\be
Q= \frac{\epsilon_{\mu \nu \rho \sigma} \Delta_{\mu \nu} \Delta_{\rho
\sigma}}{8 N N_1 N_2}= \frac{\mathrm{Pf}(\Delta)}{N N_1 N_2}
\ee
where $\mathrm{Pf}(\Delta)$ is the Pfaffian of the antisymmetric
matrix $\Delta_{\alpha \beta}$.
Although computed with the use of the connection $\hat{A}$ the
topological charge only depends on the transition matrices
$\Omega_\alpha$.

Notice that the constant field strength connection $\hat A$, being proportional to the
single Lie algebra generator $T$, is essentially
abelian, but the bundle is non-abelian. These constant abelian gauge
fields  are  solutions  of the classical equations of motion (for  constant metric
tensor), but  are in general unstable. Self-dual
solutions are obviously stable. 

We can now use symmetries of the system to write the solution in  a
simpler form. First of all we use the  freedom to redefine the basis
of the lattice $\Lambda$. This can be done by means of
$SL(4,\mathbb{Z})$ transformations. By well-known
properties~\cite{peniscola} we can find an appropriate basis such that only
$\Delta_{ 0 3}=-\Delta_{ 3 0 } \equiv \Delta_A$ and $\Delta_{ 1
2}=-\Delta_{ 2 1 } \equiv \Delta_B$ are non-zero. The topological
charge is now simplified to
\be
Q= \frac{\Delta_A \Delta_B}{N N_1 N_2}
\ee
It is easy to show using Eqs.~\eqref{conds} that $\Delta_A \Delta_B$
is proportional to $N_1 N_2$ and the topological charge has the form
put forward  by `t Hooft.

If we want to find solutions having minimum non-zero action we should
take $\Delta_A \Delta_B=N_1 N_2$. Thus, a general solution is provided
by introducing 4 positive integers $M_{A 1}$, $M_{A 2}$, $M_{B 1}$ and $M_{B 2}$
and writing 
\be 
\Delta_A =M_{A 1} M_{A 2}\ ; \quad   \Delta_B=M_{B 1} M_{B 2}\ ;
\quad N_1=M_{A 1} M_{B 1}\ ; \quad   N_2=M_{A 2} M_{B 2}\ 
\ee
`t Hooft made the special choice $M_{A 2}=M_{B 1}= 1$.

Now we will enforce self-duality. The explicit formulas  do  depend 
on the metric. Essentially, the relevant piece of information needed
is the value of the scalar products of the  basis vectors of our lattice $\Lambda$:
\be
\hat g_{\alpha \beta}=(e_\alpha, e_\beta)
\ee
where we use the notation $(\cdot, \cdot)$ for the scalar product. 
This information translates into the lengths of $e_\alpha$, the areas
of the $\alpha-\beta$ faces, the total volume of the torus, etc. 
In retrospective, we can say that the idea of `t Hooft was to choose the
metric $g=\hat{g}$ in such a way as to enforce the self-duality condition for the constant
field-strength connection.  We will explain this better in the next
paragraphs.

Although, we have restricted ourselves to a flat metric, we will still
need to use different sets of coordinates related by linear
transformations. In a given set of coordinates, the expression of the
metric $ds^2= g_{\mu \nu} dx^\mu dx^\nu$ defines a specific constant
matrix $g_{\mu \nu}$. The self-duality condition expressed in this
coordinate system is given by 
\be
F_{\mu \nu}=\frac{\sqrt{\det(g)}}{2} \epsilon_{\alpha \beta \mu \nu}
F^{\alpha \beta} \equiv \tilde F_{\mu \nu}
\ee
where the metric tensor $g_{\mu \nu}$ and its inverse  $g^{\mu \nu}$ are 
used for  lowering and rising  indices in the standard way. In our
particular problem  there are two natural systems of coordinates which
will be useful.  The first one is that in which the coordinates are
aligned along the basis vector of the lattice: $x=\sum_\alpha e_\alpha
y^\alpha$. These coordinates (unit-period coordinates) have the advantage 
that the torus has periods of 1 in each direction ($y^\mu
\longrightarrow y^\mu +1$) . It is also in this coordinate system in
which we can write down easily the form of the constant field strength tensor:
\be
\label{eqF}
\hat{F}= \frac{2 \pi}{N} ( \Delta_A dy^0\wedge dy^3 + \Delta_B
dy^1\wedge dy^2 )  T \equiv \frac{N_1 N_2}{2 N} T\,  f_{\alpha \beta}
\, dy^\alpha \wedge 
d y^\beta 
\ee
The metric in these coordinates can be written as $ds^2= \hat g_{\alpha
\beta} dy^\alpha dy^\beta$. This gives the lengths of the basis vectors
$\| e_\alpha \| =\sqrt{\hat g_{\alpha \alpha}}$  and the volume of the torus  
$\V =\sqrt{\det(\hat g)}$.

The other quite natural coordinate system is the one in which the
metric tensor is the unit matrix. 
We label the corresponding coordinates by $z^a$. The  change
of variables is produced by the vierbein $V_\alpha^a$. 
(which in our case is just a constant matrix):
\be
z^a = V_\alpha^a y^\alpha \ ; \quad y^\alpha= W^\alpha_a z^a
\ee
where 
\be
\sum_a  V_\alpha^a V_\beta^a= \hat g_{\alpha \beta} \  ;  \quad \sum_a V_\mu^a
W^\nu_a = g^\mu_\nu
\ee
These conditions do not specify the $z^a$ coordinates uniquely. We are
still free to perform  orthogonal transformations in the $z^a$ variables.
This freedom can be used to adopt a canonical form for the field
strength in these coordinates $\hat F=\frac{1}{2} T \bar{F}_{a b} dz^a\wedge
dz^b$ with 
\be
\bar{F} = \frac{2 \pi}{N} \begin{pmatrix} 0 & 0 & 0 & \f_A \cr 0 & 0
& \f_B &0 \cr
0 & -\f_B & 0 & 0 \cr
-\f_A & 0 & 0 & 0 \end{pmatrix}
\ee
with $\f_A \ge \f_B > 0$. The quantities $\f_A$ and $\f_B$ can be
expressed in terms of coordinate invariant quantities. In particular,
we can take $\rho_S\equiv \frac{1}{4} \Tr(F_{\mu \nu} F^{\mu \nu})$ and 
$\rho_Q \equiv \frac{1}{4} \Tr(F_{\mu \nu} \tilde F^{\mu \nu})$. The
formula is 
\bea
\f_A= \frac{\sqrt{N N_1 N_2}}{2 \pi}  \sqrt{\rho_S + \sqrt{\rho_S^2-\rho_Q^2}} \\
\f_B= \frac{\sqrt{N N_1 N_2}}{2 \pi} \sqrt{\rho_S - \sqrt{\rho_S^2-\rho_Q^2}} 
\eea
The self-duality condition is then simply given by 
$\f_A=\f_B$.
Notice that $\rho_Q= 4\pi^2 Q/\V = 4 \pi^2 \f_A \f_B/(N N_1 N_2)=  4 \pi^2 \Delta_A
\Delta_B/ (N N_1 N_2 \V)$, where $\V$ is the volume of the torus
(another invariant). Thus, in  the self-dual case
$\f_A=\f_B=\sqrt{N_1 N_2 /\V}$. 

There is a whole family of metrics $g=\hat g$ for which the
constant field strength connection is self-dual. For any of these
cases we have a minimum action  fractional instanton with constant field strength.
A particularly simple case is the one chosen by `t Hooft,  in which the metric tensor is diagonal in the unit-period
coordinates:
\be
ds^2= \sum_\mu l_\mu^2 dy^\mu dy^\mu
\ee
This amounts to assuming that the generators of the lattice $\Lambda$
are orthogonal and have length $||e_\alpha||=l_\alpha$. 
Then we have
\be
\rho_Q= \frac{4 \pi^2}{N N_1 N_2} \frac{\Delta_A \Delta_B}{l_0 l_1 l_2 l_3} \ ; \quad \rho_S=
\frac{4 \pi^2}{N N_1 N_2}  \left(\frac{\Delta_A^2}{2 l_0^2 l_3^2}+
\frac{\Delta_B^2}{2 l_1^2 l_2^2}\right) 
\ee
giving $\f_A=\Delta_A/(l_0 l_3)$ and  $\f_B=\Delta_B/(l_1 l_2)$. The self-duality condition then becomes
\be
\frac{\Delta_A}{l_0 l_3}= \frac{\Delta_B}{l_1 l_2}
\ee
Thus, the ratios of areas of the two twisted planes must be a
particular rational number. Notice that, even within the set of diagonal
matrices, there are many solutions since, for example,  multiplying $l_0$ by any
number and dividing $l_3$ by the same number does not alter the
self-duality. 

It is possible to obtain the most general constant symmetric matrix
$\hat g_0$ for which
the constant field strength connection is self-dual. For that purpose
we realize that given  an antisymmetric matrix $X$, one has 
\be
 \frac{1}{2} \epsilon_{\mu \nu \rho \sigma}
X_{\rho \sigma} = - \mathrm{Pf}(X) (X^{-1})_{\mu \nu}
\ee
Hence, if we apply this expression to the antisymmetric tensor $f_{\mu
\nu}=\Tr(T \hat F_{\mu \nu})$ defined in Eq.~\eqref{eqF}  we obtain 
\be
\tilde f_{\mu \nu}= -\frac{\mathrm{Pf}(f)}{\sqrt{\det \hat g_0}} (\hat
g_0 f^{-1}\hat g_0)_{\mu
\nu}
\ee
From here we see that the there is no constraint on the determinant of
$\hat g_0$ (conformal invariance), and on the value of the Pfaffian of $f$. 
If we define $J=\hat g_0^{-1}f\cdot(\det(\hat g_0)^{1/4}/\sqrt(\mathrm{Pf}(f))$,
then the self-duality condition becomes $J^2=-1$. This defines an
almost complex structure. Hence, $f$, $J$ and $\hat g_0$ are a
compatible triplet. Self-duality is achieved for all metrics of the
form 
\be
\label{metricsolution}
 \hat g_0= - f J 
\ee
up an arbitrary multiplicative constant. The compatibility condition 
ensuring that the matrix $\hat g_0$ is symmetric reads 
\be
f=J^t f J
\ee
which expresses the fact that $J$ is an element of the symplectic
group $Sp(4,\mathbb{R})$. Thus, given any element of the group and
inserting it in Eq.~\ref{metricsolution}, we get all the constant
metric tensors for which the constant field strength is self-dual.

Concerning the choice of integers $M_{A 1}$,  $M_{A 2}$,  $M_{B 1}$,
$M_{B 2}$, it is convenient to restrict ourselves to $N_1$ and $N_2$ being
coprime. Otherwise, $\gcd(N_1,N_2)$ divides also $N$, $\Delta_A$ and
$\Delta_B$,  and by dividing by this greater common divisor we can
reduce the problem to this case.  With the coprime condition we can easily solve for all
quantities. We conclude that $k^{(a)}=M_{A a} \hat{k}^{(a)}$  with 
$\hat{k}^{(a)}$  coprime with $M_{B a}$, satisfying
\be
\hat{k}^{(2)}M_{B 1}- \hat{k}^{(1)}M_{B 2}=1
\ee
 In an analogous fashion 
$m^{(a)}=M_{B a} \hat{m}^{(a)}$,  with
$\hat{m}^{(a)}$  coprime with $M_{A a}$. These two conditions imply
that the matrices $\Gamma^{(a)}_0$ and $\Gamma^{(a)}_3$  generate 
an $M_{B a}^2$ dimensional irreducible algebra, while  $\Gamma^{(a)}_1$ and
$\Gamma^{(a)}_2$ generate an $M_{A a}^2$ dimensional algebra. In other words we
can write our twist-eating  matrices as tensor products
\bea
\Gamma^{(a)}_0&=&\hat\Gamma^{(a)}_0 \otimes \II_{M_{A a}}\ ; \quad \quad
\Gamma^{(a)}_3=\hat\Gamma^{(a)}_3 \otimes \II_{M_{A a}}\  \\
\Gamma^{(a)}_1&=& \II_{M_{B a}} \otimes \hat\Gamma^{(a)}_1 \ ; \quad \quad
\Gamma^{(a)}_2= \II_{M_{B a}} \otimes \hat\Gamma^{(a)}_2  
\eea
where $\hat\Gamma^{(a)}_{0,3}$ are $M_{B a}\times M_{B a}$ matrices 
satisfying 
\be
\hat\Gamma^{(a)}_0 \hat\Gamma^{(a)}_3= e^{2 \pi i \hat{k}^{(a)}/M_{B
a}} \hat\Gamma^{(a)}_3 \hat\Gamma^{(a)}_0 
\ee
and $\hat{k}^{(a)}=k^{(a)}/M_{A a}$. A similar relation follows for
$\hat\Gamma^{(a)}_{1,2}$ , replacing $\hat{k}$ by $\hat{m}$ and $M_{B
a}$ by $M_{A a}$.

More specifically, we can choose a basis in which $\hat\Gamma^{(a)}_0$ and 
$\hat\Gamma^{(a)}_1$ are diagonal. Then we can express all matrices $\hat\Gamma$  in
terms of t Hooft clock matrices ($s\in\{0,1,\ldots,N-1\}$):
\bea
(Q_N)_{s s'}&=& e^{i \theta_N}  e^{2 \pi i s /N} \delta(s-s')\\
(P_N)_{s s'}&=& e^{i \theta_N} \delta(s+1-s')
\eea
where $\theta_N=$ vanishes for odd $N$ and equals $\pi/N$ for even $N$
to ensure that the matrices belong to SU(N). 
Now we can write
\bea
\nonumber
\hat \Gamma_0^{(a)}=Q_{M_{B a}} \ &;& \quad \hat \Gamma_1^{(a)}=Q_{M_{A a}} \\
\hat \Gamma_3^{(a)}=(P_{M_{B a}})^{-\hat k^{(a)}} \ &;& \quad \hat
\Gamma_2^{(a)}=(P_{M_{A a}})^{-\hat m^{(a)}}
\eea
 This means that all the basis vectors  of the $N_a$ dimensional space
are labelled by a pair of integers $(s_{A a}, s_{B a})$, with $0\le
s_{X,a}\le M_{X a}-1$. We will be using this basis in what follows.

We emphasize that our construction generates all constant field
strength fractional instantons. This includes  the SU(2) case dealt in 
Ref.~\cite{su2paper}, as well as the apparently different looking
solutions appearing in Ref.~\cite{Montero:2000mv}. 
It is convenient to rewrite
the main equations in matrix form as follows:
\be
\mathbf{\hat N} \mathbf{M}\equiv \begin{pmatrix} \hat k^{(2)} &
\hat k^{(1)}  \cr  -\hat m^{(1)} & \hat m^{(2)} \end{pmatrix} 
 \begin{pmatrix} M_{B1} & M_{A2}
   \cr  -M_{B2} & M_{A 1} \end{pmatrix} = 
   \begin{pmatrix} 1& k \cr  -m & 1  \end{pmatrix}    
\ee
where the matrix $\mathbf{M}$ has determinant equal to $N$, and the
matrix $\mathbf{\hat N}$ has determinant $(km+1)/N$. As a more complex
example one might take $N=43$ and split it into $N_1=15$, $N_2=28$.
This gives integers  $M_{B 1}=3$ $M_{B 2}=4$ $M_{A 1}=5$
$M_{A 2}=7$, and hence  $\Delta_A=35$, $\Delta_B=12$. 
This gives $\hat{k}^{(1)}=2$, $\hat{k}^{(2)}=3$, $\hat{m}^{(1)}=2$,
$\hat{m}^{2}=3$ and hence $k=31$ and $m=18$. 

A very interesting example is provided by the case in which $N_1=M_{A
1}$ and $N_2=M_{B2}$ are two successive Fibonacci numbers. Then $N$
becomes the next number in the sequence. As in other related
problems~\cite{Chamizo:2016msz}, running over the index of the Fibonacci
sequence  defines a nice way  to take the large $N$ limit,  in
which the field strength tends to a finite value. Furthermore, the
corresponding ratio of areas for the self-duality condition 
is given by the golden ratio. This and other possible choices involving 
generalized Fibonacci sequences are worth of being explored in greater
detail. 

\section{Deforming constant field strength  connections}
\label{deformation}
Here we will address the case in which the constant field strength
connection is not self-dual. Our strategy will be to construct the 
non-constant self-dual connection by deforming
the previous constant connections obtained in the previous section.

Any vector potential  defined on the bundles considered can be written as  
\be
A_\mu(x)= \hat{A}_\mu + \delta_\mu(x) 
\ee
where $\hat{A}_\mu$ is the  constant field strength associated
to the transition matrices. The main advantage is that $\delta_\mu$
transform homogeneously under translations by the generators of the
lattice $\Lambda$. To express the twisted boundary conditions  it is
convenient to split $\delta_\mu$  into the $N_1$ and $N_2$ rows and columns:
\be
\delta_\mu(x) = \begin{pmatrix} S_\mu^{(1)}(x) & \W_\mu(x) \cr
\W_\mu^\dagger(x) &
S_\mu^{(2)}(x) \end{pmatrix}
\ee
Thus, $S_\mu^{(a)}(x)$ is an $N_a\times N_a$ hermitian matrix satisfying
\be
\label{Sboundary}
S_\mu^{(a)}(x+e_\alpha)= \Gamma_\alpha^{(a)} S_\mu^{(a)}(x)
\Gamma_\alpha^{\dagger (a)}
\ee
On the other hand the $N_1\times N_2$ matrix $\W_\mu$ satisfies
\be
\label{Wboundary}
\W_\mu(x+e_\alpha)=\exp\{i \pi N  \hat{w}(e_\alpha,x) /(N_1 N_2)\}
\ \Gamma_\alpha^{(1)} \W_\mu(x) \Gamma_\alpha^{\dagger
(2)}
\ee
Following the choices done in the previous section, we will take the
$e_\alpha$ that brings  $\hat{w}(e_\alpha,e_\beta)$ to canonical form.

Now we can compute the field strength
\be
F_{\mu \nu}(x) = \hat{F}_{\mu \nu} + \hat{D}_\mu \delta_\nu -
\hat{D}_\nu\delta_\mu -i [\delta_\mu, \delta_\nu]
\ee
The operators $\hat{D}_\mu$ are the
covariant derivatives (in the adjoint representation) with respect to
the constant field strength connection $\hat A$. 
The field tensor can also be decomposed into blocks 
\be
F_{\mu \nu}(x) =  \begin{pmatrix} F^{(1)}_{\mu \nu}(x) &
\mathcal{F}_{\mu \nu} (x) \cr
\mathcal{F}^\dagger_{\mu \nu}(x) & F^{(2)}_{\mu \nu}(x)
\end{pmatrix}
 \ee
with coefficients   given by 
\bea
\nonumber
F^{(1)}_{\mu \nu}(x) &=& \hat{F}^{(1)}_{\mu \nu} + \partial_\mu
S_\nu^{(1)}(x) - \partial_\nu S_\mu^{(1)}(x) -i [ S_\mu^{(1)},
S_\nu^{(1)} ] -i \W_\mu \W_\nu^\dagger +i  \W_\nu \W_\mu^\dagger \\
\label{fieldstrength}
F^{(2)}_{\mu \nu}(x) &=& \hat{F}^{(2)}_{\mu \nu} + \partial_\mu
S_\nu^{(2)}(x) - \partial_\nu S_\mu^{(2)}(x) -i [ S_\mu^{(2)},
S_\nu^{(2)} ] -i  \W_\mu^\dagger \W_\nu+i   \W_\nu^\dagger \W_\mu \\
\nonumber
\mathcal{F}_{\mu \nu} (x)  &=& \bar{D}_\mu \W_\nu- \bar{D}_\nu \W_\mu
- i S_\mu^{(1)} \W_\nu +i S_\nu^{(1)} \W_\mu -i \W_\mu S_\nu^{(2)} +i 
 \W_\nu S_\mu^{(2)}
\eea
where $\bar{D}_\mu$ is the covariant derivative with respect to a U(1)
gauge field 
whose constant field strength $f_{\mu \nu}=\Tr(T \hat F_{\mu \nu})$ was defined in
Eq.~\eqref{eqF}.

Up to now everything is independent on the metric and hence on the
choice of coordinates. The  self-duality condition can be expressed by
setting to zero the projection onto the self-dual part. This can be
written as follows
\be
\frac{1}{2}\bar\eta^{\mu \nu}_i F_{\mu \nu}=0
\ee
where $\bar\eta^{\mu \nu}_i$ for $i=1,2,3$ are a basis of the antiself-dual  
tensors. For unit metric tensor they coincide with the symbols
$\bar\eta^{a b}_i$ introduced by `t Hooft. For the unit period  metric they can be
written as 
\be
\bar\eta^{\mu \nu}_i= W^\mu_a W^\nu_b \bar\eta^{a b}_i
\ee

The contribution of the constant
field strength is then 
\be
\frac{1}{2}\bar\eta^{\mu \nu}_i \hat F_{\mu \nu}= \delta_{i 3 } (\hat F_{0 3} -\hat
F_{1 2}) = \frac{2 \pi}{N} (\f_A-\f_B) \delta_{i 3} T
\ee
which vanishes in the self-dual case.  The strategy put forward in our
paper~\cite{su2paper} is to treat the  difference $\epsilon\equiv
(\f_A-\f_B)$ 
%\be
%\epsilon \equiv \frac{\Delta_A}{l_0 l_3}- \frac{\Delta_B}{l_1 l_2}
%\ee
as an expansion parameter and compute the self-dual connection as a
power series expansion in this parameter. When only a few orders are
computed the approximation becomes closer to the exact
result the smaller the value of $\epsilon$. Indeed, this was verified
in Ref.~\cite{su2paper},  for the SU(2) case with the  diagonal metric,  by computing the analytic expressions and
comparing them with the numerical solution obtained by a minimization
method. The solution now has a lumpy structure with a peak in the
action density  at a
particular point. Obviously the 4 moduli parameters are associated
with the space-time coordinates of the peak.

In what follows we will extend the previous construction to SU(N). For
that purpose it is important to revise  the details of the method as
it appears  for  SU(N) case.

The first observation is that the off-block part of the deformation
$\W_\mu$ becomes a power series in odd powers of $\sqrt{\epsilon}$:
\be
\W_\mu = \sqrt{\epsilon}\, \sum_{n=0}^\infty \epsilon^n \W_{\mu,  n}(x)
\ee
On the other hand the block terms $S_\mu^{(a)}$ become power series in
$\epsilon$ starting at order 1:
\be
S_\mu^{(a)} = \epsilon  \sum_{n=0}^\infty \epsilon^n  S_{\mu , n}^{(a)}
\ee
The even or odd powers of  $\sqrt{\epsilon}$ apply to $F^{(a)}$ and
$\mathcal{F}$ as well, as can be seen from the
expression~\eqref{fieldstrength}.

Now, before going into the actual calculation of the coefficients 
$S_{\mu , n}^{(a)}$ and $W_{\mu,  n}(x)$, let us explain how the first
few terms in the expansion proceed, because that clarifies the general
procedure with certain subtleties involved. The first term in the
expansion of the self-dual part of the action is actually of order
$\sqrt{\epsilon}$:
\be
\label{firsteq}
0=\bar{\eta}^{\mu \nu}_i \bar D_\mu \W_{\nu,  0} 
\ee
This is an homogeneous equation so that the solution is only fixed up
to a multiplicative constant. Thus, it is unclear to what extent is the
contribution of order $\sqrt{\epsilon}$. This becomes clear when
looking at the equation at order $\epsilon$:
\bea
\label{secondeq}
0=\frac{2 \pi}{N}  T \delta_{i 3}+ \frac{1}{2} \bar{\eta}^{\mu \nu}_i
\begin{pmatrix} \partial_\mu
S_{\nu,0}^{(1)}(x) - \partial_\nu S_{\mu, 0}^{(1)}(x)   & 0 \cr 
0 &  \partial_\mu S_{\nu,0}^{(2)}(x) - \partial_\nu S_{\mu, 0}^{(2)}(x)  
\end{pmatrix} +
 \\ \nonumber
+ \frac{1}{2} \bar{\eta}^{\mu \nu}_i
\begin{pmatrix}   -i \W_{\mu,0}
\W_{\nu,0}^\dagger +i  \W_{\nu, 0} \W_{\mu, 0}^\dagger & 0 \cr
0 &   -i  \W_{\mu, 0}^\dagger W_{\nu, 0} +i   \W_{\nu, 0} ^\dagger W_{\mu, 0}
 \end{pmatrix}
\eea
The first term comes from the constant field strength part, which as
we saw before is of order $\epsilon$. Now if we integrate this
equation over the torus, the term containing derivatives vanishes and
we get 
\be
\label{norm}
0=\frac{2 \pi \V}{N} T  \delta_{i 3} - \frac{i}{2} \bar{\eta}^{\mu
\nu}_i \int_\mathcal{T} dx \begin{pmatrix} \W_{\mu,0}
\W_{\nu,0}^\dagger -  \W_{\nu, 0} \W_{\mu, 0}^\dagger & 0 \cr
0 &  \W_{\mu, 0}^\dagger \W_{\nu, 0} -  \W_{\nu, 0} ^\dagger \W_{\mu, 0}
\end{pmatrix}
\ee
where $\V$ is the volume of the torus. Indeed, one can multiply the
equation by the generator $T$ and take the trace to obtain
\be
0=\frac{2 \pi \V}{N}   \delta_{i 3}  - i \bar{\eta}^{\mu
\nu}_i \int_\mathcal{T} dx \Tr(\W_\mu \W_\nu^\dagger)
\ee
It is now obvious that this equation fixes the normalization of $
\W_{\mu, 0}$ up to a phase.

The arbitrarity of the phase can be put into a wider context by
investigating  the multiplicity of solutions.
Obviously symmetries imply that the solution is non-unique.
First of all, one has gauge transformations. As in our previous paper
we fix them by imposing the background field gauge $\hat D_\mu
\delta_\mu=0$, leading to 
\be
\partial_\mu S_{\mu}^{(a)}=0 \ ; \quad  \bar{D}_\mu \W_\mu = 0
\ee
There is a remaining invariance under those global gauge
transformations which are consistent with the boundary conditions.
Indeed, this freedom is connected to the phase arbitrarity of $\W$.

Apart from the phase arbitrarity  notice that in Eq.~\eqref{secondeq}  $S^{(a)}_\mu$ only
enters through its derivative. Thus, one can always add a constant,
which because of the boundary conditions must be proportional to the
identity in each block. The traceless condition then fixes this to be a constant
times the generator $T$. Finally, one realizes that the arbitrarity
can be associated to space-time translation, being equivalent to  a
shift $x\longrightarrow x-x_0$ in the original spatial constant
solution. One can add a condition to fix this arbitrarity and obtain a
unique solution.  This is very similar to the discussion and procedure 
employed in Ref.~\cite{su2paper} when dealing with the SU(2) case.

After this explanation we proceed to the actual calculation to first
order in $\epsilon$ which is done in the next section.
\section{Non-constant fractional intanton to order $\epsilon$}
\label{orderepsilon}
In this section we present the calculation up to order $\epsilon$, as
was done in  our previous paper for SU(2). The calculation will be
split into two subsections leading with $\W_{\mu, 0}$ and
$S^{(a)}_{\mu ,  0}$ respectively.

\subsection{The first equation}
The first part of the calculation involves  the determination of
$\W_{\mu, 0}$. This function satisfies the boundary conditions
Eq.~\eqref{Wboundary} and the equation 
\be
\frac{1}{2} \bar\eta^{\mu \nu}_i \bar{D_\mu} \W_{\nu, 0}=0
\ee
where $\bar D_\mu$ is the covariant derivative with respect to the
abelian gauge field with  constant field strength $f=\Tr(T\hat F)$. 
In unit-period coordinates $y^\mu$ and orthonormal
coordinates $z^a$ we can write
\be
\label{f_exp}
f=  \frac{ 2 \pi }{\Delta_B} dy^0\wedge dy^3 +  \frac{
2 \pi }{\Delta_A} dy^1\wedge dy^2\ = \frac{ 2 \pi \f_A}{N_1
N_2} dz^0\wedge dz^3 +  \frac{ 2 \pi \f_B}{N_1
N_2} dz^1\wedge dz^2
\ee

 We can restate the boundary conditions
by introducing operators $\Op_\alpha$ as follows:
\be
\Op_\alpha = e^{-i  f_{\mu \nu} e^\mu_\alpha x^\nu/2} \, \del_\alpha 
\ee
where $\del_\alpha$ is the operator that shifts $x$ by $e_\alpha$:
\be
\del_\alpha \Psi(x) = \Psi(x+e_\alpha)
\ee
The operators satisfy the  relations 
\be
\Op_\alpha \Op_\beta = e^{2 \pi i \Delta_{\alpha \beta}/(N_1 N_2)}
\Op_\beta \Op_\alpha 
\ee
Now the boundary conditions can be rewritten as 
\be
\Op_\alpha  \W = \Gamma^{(1)}_\alpha \W \Gamma^{(2)\dagger}_\alpha
\ee

Notice that the operators $\Op_{0,3}$ commute with $\Op_{1,2}$. We have
\bea
\label{algebra}
\Op_0 \Op_3=  e^{2 \pi i /\Delta_B} \Op_3 \Op_0 \\
\Op_1 \Op_2=  e^{2 \pi i /\Delta_A} \Op_2 \Op_1 
\eea
Now we can simultaneously diagonalize $\Op_1$ and $\Op_0$ which are
unitary operators. Through the boundary conditions this is equivalent
to diagonalizing $\Gamma^{(a)}_0$ and $\Gamma^{(a)}_1$. This is the
same as going to the basis that was presented in
section~\ref{constant}. It is convenient to label the matrix elements 
of $\W(x)$  in terms of two indices $l_B=(s_{B 1} M_{B 2} - s_{B 2} M_{B
1})$ and $l_A=(s_{A 1} M_{A 2} - s_{A 2} M_{A 1})$. In this notation
the function $\W_{l_A l_B}(x)$ satisfies the following boundary
conditions
\bea
\Op_0 (\W)_{l_A\, l_B} &=& e^{i\theta_B} e^{2\pi i l_B/\Delta_B}
(\W)_{l_A\, l_B}\  ; \quad
\Op_3 (\W)_{l_A\, l_B} =  e^{i\theta_B} (\W)_{l_A\, l_B+1} \\
\Op_1 (\W)_{l_A\, l_B} &=&  e^{i\theta_A} e^{2\pi i l_A/\Delta_A}
(\W)_{l_A\, l_B}\   ; \quad  \Op_2
(\W)_{l_A\, l_B} = e^{i\theta_A} (\W)_{l_A+1\, l_B} 
\eea 
where $\theta_{A,B}$ is zero if $\Delta_{A,B}$ is odd. In general, we
have $\theta_X=\pi \epsilon_X/\Delta_X$ with $\epsilon_X\equiv M_{X1}-M_{X 2} \bmod 2$.
These boundary conditions imply that once $ \bar{\W}(x)\equiv (\W)_{0 0}(x)$ is known, we
can immediately solve for $(\W)_{l_A\, l_B}$ as follows
\be
\label{full_sol}
(\W)_{l_A\, l_B}(x)= (\Op_3)^{l_B} (\Op_2)^{l_A} \bW (x) =
e^{i \pi (y^0 l_B/\Delta_B + y^1 l_A/\Delta_A)} \bW(x+l_B e_3
+l_A e_2)
\ee
We recall that the function $\bW$ satisfies the following boundary
conditions 
\bea
\Op_0 \bW = e^{i \theta_B}  \bW\ &;& \quad (\Op_3)^{\Delta_B}
\bW = e^{i
\theta_B\Delta_B}  \bW \\
\Op_1 \bW = e^{i \theta_A}  \bW\ &;& \quad (\Op_2)^{\Delta_A}
\bW = e^{i
\theta_A\Delta_A}  \bW
\eea

Now notice that all operators $\bar{D}_\mu$ commute with $\Op_\alpha$
and furthermore they are scalar and do not mix different components of
$\W$.  All the problem reduces to that of an
abelian connection.
In particular, our first equation reduces to  a scalar equation
involving only $\bW_\mu(x)$:
\be
\frac{1}{2} \bar\eta^{\mu \nu}_i \bar{D}_\mu \bW_{\nu, 0} =0
\ee 
In the $z$ coordinate system the projection operator becomes just `t
Hooft symbol, so that using the same strategy as in our SU(2) paper,
we can reformulate the problem by introducing $2\times 2$ matrices
$\sigma_a=(\II_2, -i \vec{\tau})$ and $\bar \sigma_a=\sigma^\dagger_a=(\II_2, i
\vec{\tau})$, where $\tau_i$ are the Pauli matrices and $\II_2$ the
$2\times 2$ identity matrix. These matrices
verify
\be
\bar \sigma_a \sigma_b = \bar\eta^{a b}_c \sigma_c
\ee
where $\bar\eta^{a b}_0=\delta_{a b}$. Now we can rewrite the equation
as 
\be
(\bar D_a \bar\sigma_a) (\bW_b \sigma_b)=0
\ee
where each of the parenthesis involves a $2 \times 2$ matrix. We have
actually added one equation which expresses the condition of
background field gauge  $\bar D_a \bW_a=0$. Being a matrix equation, the
previous condition imposes 4 real equations. Just as for the SU(2) case 
the equation has a solution when $(\bW_b \sigma_b)$ consist only of
the 11 element. This occurs for $\bW_1=\bW_2=0$ and $\bW_3=i\bW_0$. In
that case, the matrix equation reduces just to 2 complex equations:
\bea
\label{EQDB}
D_B \bW_0 \equiv (\bar D_0+i \bar D_3)\bW_0=0 \ &\Leftrightarrow& \ (\frac{\partial}{\partial
z^0}+i \frac{\partial}{\partial z^3}) \bW_0= -\frac{\pi \f_A}{N_1 N_2}
(z^0+i z^3) \bW_0 \\
\label{EQDA}
D_A \bW_0 \equiv (\bar D_1+i \bar D_2)\bW_0=0  \ &\Leftrightarrow& \ (\frac{\partial}{\partial
z^1}+i \frac{\partial}{\partial z^2}) \bW_0= -\frac{\pi \f_B}{N_1 N_2}
(z^2+i z^2) \bW_0
\eea
The choice of $\bW_a$ is justified precisely to keep only these two
conditions. Why precisely these two is clear from our previous work on
the subject~\cite{su2paper,GonzalezArroyo:2004xu} and will be explained below.  
The equations are essentially two copies of the equations involved in 
the Bradlow expansion for vortices on the
2-torus. The treatment  performed in Ref.~\cite{GonzalezArroyo:2004xu} 
is to write the equations in terms of complex coordinates, which fixes
the solution up to a  holomorphic function. The latter is fixed by the
boundary conditions.  In the two dimensional  case these boundary conditions
led to  the Jacobi theta functions and those with rational
characteristics. In our case, something very similar follows for the
case in which $\hat g$ (the metric tensor in  unit-period coordinates)
is diagonal. 

In treating the general case, we consider more instructive to follow an
alternative method which is more constructive. For that purpose we
express eqs.~\eqref{EQDB}-\eqref{EQDA} and the boundary conditions 
in terms of the unit-period coordinates $y$. The two equations can be written as 
\be 
U_X^\alpha ( \frac{\partial}{\partial y^\alpha} + i f_{\alpha
\beta}y^\beta/2 ) \bW_0 =0
\ee
where $X\in \{A,B\}$,  $U_A=W_0+iW_3$ and $U_B=W_1+i W_2$. We remind
the reader that the vectors $W_a^\alpha$ are
the inverse of  the vierbein, which in these coordinates coincide with
the lattice generators $e_\alpha^a$. Given the form of the equation we
will try a solution which is the exponential of a quadratic form 
\be
\WZERO= \exp\{-\half y^\alpha y^\beta R_{\alpha \beta}\}
\ee
Obviously, the matrix $R$ is symmetric. Applying the previous equation
to our ansatz we get 
\be
\label{matrix_eq}
U_A^\alpha \left( - R_{\alpha \beta} +  \frac{i}{2} f_{\alpha
\beta}\right )y^\beta  =0 
\ee
This equation alone does not fix the matrix $R$ uniquely. Now we should 
impose the boundary conditions. 

We first impose the boundary conditions with respect to translations
by $e_0$ and $e_1$. This demands that 
\be
-R_{\alpha \beta} -\frac{i}{2} f_{\alpha \beta}=0
\ee
valid for $\alpha=0,1$ and $\beta$ arbitrary. Given the symmetry of $R$ 
this equation fixes the matrix $R$ up to the $2\times 2$
submatrix $\bar{R}$
with $\alpha ,\beta\in\{3,2\}$. The next step is to return to
Eq.~\eqref{matrix_eq} with the information that we have obtained on
the structure of $R$. The best way to obtain the solution is by
expressing the equation in terms of $2\times 2$ matrices. We write
$U_{11}$ for the $2\times 2$ matrix with components $U_X^\alpha$ with
$\alpha=0,1$. We call $U_{12}$ the corresponding one matrix  for
$\alpha=3,2$. Now setting $F_2=2\pi\,  \mathrm{diag}(1/\Delta_B,
1/\Delta_A)$ 
 we can write 
\be
i U_{1 1} F_2 - U_{1 2} \bar{R} =0
\ee
This allows us to solve for $\bar{R}$
\be
\bar{R} =i U_{1 2}^{-1} U_{1 1} F_2
\ee
To see the consistency of the solution we should still verify that the
so obtained matrix is symmetric. This can be deduced from the form of
$f$ in both coordinate systems (Eq.~\eqref{f_exp}). We leave the
verification to the reader. 

We now summarise the form of $R$:
\be
R_{\alpha \beta} y^\alpha y^\beta=-\frac{2 i \pi}{\Delta_B} y^0 y^3
-\frac{2 i \pi}{\Delta_A} y^1 y^2 + \bar R_{A A} (y^2)^2 +
\bar R_{B B} (y^3)^2 +\bar R_{A B} y^2 y^3 
\ee
For the diagonal metric case $\bar R_{A B}=0$, $\bar R_{A A }=\frac{2
\pi l_2}{l_1 \Delta_A}$ and  $\bar R_{B B}=\frac{2
\pi l_3}{l_0 \Delta_B}$

We have succeeded in constructing a solution of the first equation
that satisfies the right boundary conditions under translations in
$y^0$ and $y^1$, but we still have not enforced the rest of boundary
conditions. This will be done constructively. Suppose that we have a
function $\Psi(x)$ and we want to impose that it satisfies the
eigenvalue equation:
\be
(\Op_3)^k \Psi = e^{i \lambda} \Psi
\ee
This can be done by projection as follows:
\be
\Psi \longrightarrow \sum_{q \in \mathbb{Z}} (\Op_3)^{kq} e^{-i
\lambda q} \Psi
\ee
In our particular case we have just to perform the following projection
\be
\label{bwprev}
\sum_{q_B,q_A \in \mathbb{Z}}   e^{-i \pi
\epsilon_A q_A -i \pi
\epsilon_B q_B}\, 
(\Op_3)^{\Delta_B q_B}
(\Op_2)^{\Delta_A q_A} \WZERO  
\ee
We point out that $\Op_\alpha$ commute with the covariant derivative,
so that the new function is still a solution of the same equation. All
we need to do is to apply the definitions to obtain the requested
solution explicitly
\be
\label{bWfinal}
\bW_0= \mathcal{N} \WZERO \sum_{q_A,q_B \in \mathbb{Z}} e^{2 \pi i
(u_B q_B + u_A  q_A)}  e^{-\half \bar{R}_{B B}(q_B
\Delta_B)^2 -\half \bar{R}_{A A }(q_A
\Delta_A)^2 -\bar{R}_{A B} \Delta_A \Delta_B q_A q_B}
\ee
where we have introduced two complex variables as follows
\be
u_B= y_0 +i \frac{\Delta_B}{2 \pi}(\bar{R}_{B B} y_3 +\bar{R}_{B A}
y_2)-\frac{\epsilon_B}{2} \ ; \quad  u_A= y_1 +i \frac{\Delta_A}{2 \pi}(\bar{R}_{A B} y_3
+\bar{R}_{A A} y_2) -\frac{\epsilon_A}{2}
\ee
Now one easily recognizes that the sum in the expression of $\bW_0$ is
just the  Riemann theta function $\Theta(\vec{u}, \tau)$ where the
$2\times 2$ symmetric matrix $\tau$ is given by
\be
\tau_{X Y}= \frac{i}{2 \pi} \Delta_X \bar R_{X Y} \Delta_Y 
\ee
for $X,Y\in \{A,B\}$.  It can be easily proven that $\tau$ satisfies Siegel positivity condition and the
function is well defined. In the special case in which $\bar{R}_{A B}=0$,
the function factorizes into a product of Jacobi theta functions. This
is the case for the diagonal metric. 

Now plugging our solution Eq.~\ref{bWfinal} into Eq.~\eqref{full_sol} we obtain the
requested solution $\W_{0 , 0}$ up to a normalization $\mathcal{N}$
which is not fixed. As explained in the previous section, this
normalization can be fixed by the space-time integral of the second
equation. This is Eq.~\eqref{norm} of the previous section. We will
now compute it. First notice that  only $\W_{0 , 0}$ and $\W_{3 , 0}=i
\W_{0 , 0}$ are non-zero. Hence, the only non-zero trace is 
\be
\Tr(\W_{0 ,0}\W^\dagger_{3, 0})=-i \sum_{l_A,l_B}  |\bW_0(x-l_A e_2-l_B e_3)|^2
\ee
Now we can perform the integral over the torus giving 
\be
\int_{\mathcal{T}} dx \Tr(\W_{0 ,0}\W^\dagger_{3, 0}) =-i \V \int_0^1
dy^0 \int_0^1 dy^1  \int_{-\Delta_B}^0 dy^3 \int_{-\Delta_A}^0 dy^2 \
|\bW_0(y)|^2
\ee
The next step is to substitute the expression \eqref{bWfinal} and
perform the integration. The interesting thing is that the integral
over $x_0$ and $x_1$ are very simple  implying that the integers $q_A$
and $q_B$ for both factors should be the same. For the purpose of
computing the final result it is much better to go back to
Eq,~\eqref{bwprev} and realize that 
\be
(\Op_a \bW_0) (\Op_a \bW_0)^* = \del_a (|\bW_0|^2)
\ee
The sum over $q_A$ and $q_B$ has then the effect of extending the
integration over $x^2$ and $x^3$ to the full real axis. We then have
\be
\int_{\mathcal{T}} dx \Tr(\W_{0 ,0}\W^\dagger_{3, 0}) =-i \V |\mathcal{N}|^2 \int_{-\infty}^\infty dy^2
\int_{-\infty}^\infty  dy^3\ |\WZERO(y^2,y^3)|^2
\ee
The final  integral is gaussian and  gives $2\pi/\sqrt{\det((\bar R
+\bar R^*)/2)}$. 
Now introducing the result into the normalization equation we get 
\be
|\mathcal{N}|^2= \frac{\sqrt{\det((\bar R +\bar R^*)/2 )}}{2 N}
\ee
For the case of the diagonal metric the determinant of $\bar R$ can be
easily determined and the result becomes
\be
|\mathcal{N}|^2=   \frac{\pi}{N} \sqrt{\frac{l_3 l_2}{l_0 l_1 N_1 N_2}} 
\ee

\subsection{The second  equation}
Now we have to look at the equation to order $\epsilon$. This is an
equation where the unknowns are the matrices $S^{(a)}_{\mu, 0}$. The
boundary conditions on the $S^{(a)}$ are given in \eqref{Sboundary}. This
can be easily solved using a modified Fourier decomposition. There
exist a basis of matrices $\hGSa(\vec{q}_c)$
satisfying~\cite{TEK2}
\be
\Ga_\mu \hGSa(\vec{q}_c) \Gda_\mu= e^{i q_{c \mu}} \hGSa(\vec{q}_c)
\ee
where $\vec{q}_c=2 \pi (n_0/M_{B a}, n_1/M_{A a}, n_2/M_{A a},
n_3/M_{B a})$ with  $0\le n_0,n_3 \le M_{B a}-1$ and $0\le n_1,n_2 \le
M_{A a}-1$ are integers. The total number of matrices is $N_a^2$, so
that the set   represents a basis of the space of $N_a\times N_a$
matrices. It is more convenient to consider that the integers $n_\mu$
are actually defined modulo $M_{X,a}$ ($M_{A,a}$ or  $M_{B,a}$
depending on the index). In addition, one
needs a normalization condition on the basis matrices. One can take 
\be
\Tr(\hGSa(\vec{q}_c) (\hGSa(\vec{p}_c))^\dagger) =
\delta(\vec{q}_c-\vec{p}_c)
\ee
where the delta function is taken modulo the corresponding congruences.
This defines the matrices up to a phase. Explicitly one can write 
\bea
\nonumber
 (\hGSo(\vec{p}))_{s s'} =\frac{1}{N_1}
e^{i s_{B 1}M_{B 2} p_3 + i s_{A,1}M_{A,2} p_2} \delta(s'_{B 1}-s_{B
1}+ \frac{p_0 M_{B 1}}{2 \pi})   \delta(s'_{A 1}-s_{A
1}+ \frac{p_1 M_{A 1}}{2 \pi}) \\
\nonumber
 (\hGSt(\vec{p}))_{s s'} =\frac{1}{N_2}
 e^{-i s_{B 2}M_{B 1} p_3 - i s_{A,2}M_{A,1} p_2} \delta(s'_{B 2}-s_{B
 2}+ \frac{p_0 M_{B 2}}{2 \pi})   \delta(s'_{A 2}-s_{A
 2}+ \frac{p_1 M_{A 2}}{2 \pi})
\eea

Finally, we can
decompose  any matrix satisfying the boundary conditions
\eqref{Sboundary}
as 
\be
\label{modfourier}
S(x)= \sum_{p \in \Lambda_*} e^{ i p_\alpha y^\alpha} \hGSa(p_c)\, 
\hat{S}(p)
\ee
where $M_{B a} p_0/(2 \pi)$, $M_{B a} p_3/(2 \pi)$, $M_{A a} p_1/(2
\pi)$ and  $M_{A a} p_2/(2 \pi)$ run over all integers, and $p_c$ is
the corresponding vector with congruent integers. In other words, 
$(p-p_c)/(2\pi)$ is an arbitrary vector of integers. 
Now given a matrix 
satisfying the boundary conditions $S(x)$, it is possible to obtain
the Fourier coefficients by the following procedure
\be
\hat{S}(p) = \prod_\alpha \left( \int_0^1 d y^\alpha  e^{-ip _\alpha
y^\alpha}\right) \Tr\left((\hGSa(p_c))^\dagger S(x)\right) 
\ee

Now let us go to the second equation and proceed as before, by
introducing the matrices $\bar\sigma_a$ and  $\sigma_b$. Given an
arbitrary vector $v_a$ we can construct matrices $\boldsymbol{\bar
v}= v_a
\bar\sigma_a$ and $ \boldsymbol{\tilde v}= v_a
\sigma_a$. Then we can write 
\be 
\boldsymbol{\bar{\partial}}\boldsymbol{\tilde{S}}= \bar\eta^{b c}_d \frac{\partial}{\partial z^b}
S_c \sigma_d
\ee
To parameterize $\boldsymbol{\tilde S}$ we write it as
$\boldsymbol{\tilde{\partial}} \boldsymbol{\tilde{G}}$.
This is always possible if $\hat{S}(p=0)=0$, i.e. when $S_b$ has no
constant term. We then see that the equation for $S$ transforms
into an equation for $G$: 
\be
 \bar\eta^{b c}_d \frac{\partial}{\partial z^b}
 S_c \sigma_d = (\frac{\partial}{\partial z^b})^2 G_d \sigma_d
\ee
Now notice that,  given that  $\W_0$ and $\W_3$ are the only
non-zero components of the other term in the equation, this implies
that only $G_3$ could be non zero. Combining this information we
write explicitly the form of $S^{(a)}_b$  
\bea
\label{SfromG}
S^{(a)}_{0 , 0}=- \frac{\partial  G_3^{(a)}}{\partial z^3} \ &;& \quad
S^{(a)}_{3 ,0} =  \frac{\partial  G_3^{(a)}}{\partial z^0} \\
S^{(a)}_{1 ,0}= \frac{\partial  G_3^{(a)}}{\partial z^2} \ &;& \quad
S^{(a)}_{2, 0} = - \frac{\partial  G_3^{(a)}}{\partial z^1} \\
\eea
Now we can write down the equation 
\be
\label{secondequation}
\begin{pmatrix} 
\mathbf{\Delta} G_3^{(1)} & 0 \cr 0 & \mathbf{\Delta} G_3^{(2)}
\end{pmatrix}= 2 \
\begin{pmatrix} \W_{0, 0} \W_{0, 0}^\dagger & 0 \cr 0 &
\W_{0, 0}^\dagger \W_{0, 0}
\end{pmatrix}
\ee
where $\mathbf{\Delta}$ is the Laplacian operator. To solve this
equation we use the Fourier decomposition. Both sides of the equation
can be written as a Fourier sum and the equality corresponds to the
equality of the Fourier coefficients. The advantage of this procedure 
is that the Laplacian operator has a simple action on the Fourier
coefficients. 

To obtain an explicit solution we need to determine the Fourier
coefficients of the left hand side of Eq.~\eqref{secondequation}. This
can be done with our formulas:
\bea
\label{Ccoeff}
 \hat C^{(1)}(p) = 2 \prod_\alpha \left( \int_0^1 d y^\alpha  e^{-ip _\alpha
y^\alpha}\right) \Tr\left((\hGSo(p_c))^\dagger \W_{0 , 0} \W_{0 , 0}^\dagger \right) 
\\
\hat C^{(2)}(p) = 2 \prod_\alpha \left( \int_0^1 d y^\alpha  e^{-ip _\alpha
y^\alpha}\right) \Tr\left((\hGSt(p_c))^\dagger \W_{0 , 0}^\dagger
\W_{0 , 0} \right)
\eea
Finally, the coefficients of $G_3^{(a)}$ are easily obtained as follows 
\be
\hat G_3^{(a)}(q) =  \frac{\hat C^{(a)}(q)}{\|q\|^2} 
\ee
where $\|q\|^2= q_\alpha q_\beta \hat g^{\alpha \beta}$, with $\hat g$ the
metric (upper indices for the inverse metric). One has to exclude 
the coefficient for $q=0$, for which the denominator is singular. 
The value of $\hat C^{(a)}(0)$ was determined earlier and used to fix
the normalization of $\W_{0, 0}$.  Applying the derivatives 
(which is easily done in the Fourier decomposition) in
Eq.~\eqref{SfromG} we obtain the Fourier coefficients of $S^{(a)}_{\mu
,0}$ which completes the solution of the second equation. 

The only missing piece for an explicit solution is to determine the
coefficients $\hat C^{(a)}(q)$ by means of the
integrals~\eqref{Ccoeff}. This poses no fundamental problem since all
are simply Gaussian integrals. The most important thing is to
determine how to do the calculation efficiently. Let me sketch very
briefly how the calculation can be done and write down the final
result. The first part is to write $\W$ as follows
\be
(\W)_{l_A l_B}=\sum_{q_A,q_B\in \mathbb{Z}} \Op_3^{\Delta_B q_B+l_B}
\Op_2^{\Delta_A
q_A+l_A} \WZERO 
\ee
Then we can write down this expression factorizing the part which
depends on $y^0$ and $y^1$ and a part that depends only on $y^3$ and
$y^2$. In this second part it is better to keep explicitly the
$\del_3$ and $\del_2$ operators. Now one can combine the result
with that of $\W^\dagger$ and $(\hat{\Gamma}(p))^\dagger$. The main
observation is that the dependence on $y_0$ and $y_1$ of the whole integral
appears as an imaginary exponential. One can integrate on these two
variables to give a delta function equating the $q_A$ and $q_B$ coming
from $\W$ and $\W^\dagger$. The final expression takes the following
form
\be
\int_0^1 dy^3 \int_0^1 dy^2 \
\sum_{n_3, n_2 \in \mathbb{Z}}  \delta_3^{n_3}  \delta_2^{n_2}
H(y_3,y_2)
\ee
for a function $H$ to be specified below. We then realize that the sum
of the $\del$ operators simply extends the integration region to
infinity, giving
\be
\int_{-\infty}^\infty  dy^3 \int_{-\infty}^\infty  dy^2 \ H(y_3,y_2)
\ee
The function $H$ is just the exponential of a quadratic form.
Introducing the 2 component column vectors $\vec{Q}$ and
$\vec{y}\equiv(y^2,y^3)$ 
we get 
\be
 H(y_3,y_2) = \mathcal{N} \exp\{-\frac{1}{2} \vec{y}^t
(\bar{R}+ \bar{R}^*)  y + \vec{Q}\vec{y}  \}
\ee

Finally the result of the calculation is 
\bea
\hat{C}^{(1)}(p)= 2  \exp\{-\frac{i}{4 \pi}(\Delta_B p_0 p_3 +\Delta_A
p_1
 p_2)-\frac{1}{4}(\frac{\bar p_0^2 + \bar p_3^2}{\f_B}+ \frac{\bar
  p_1^2 + \bar p_2^2}{\f_A}) \} \\
\hat{C}^{(2)}(p)= -2 \exp\{\frac{i}{4 \pi}(\Delta_B p_0 p_3 +\Delta_A
p_1
    p_2)-\frac{1}{4}(\frac{\bar p_0^2 + \bar p_3^2}{\f_B}+ \frac{\bar
         p_1^2 + \bar p_2^2}{\f_A}) \}
\eea
The $\bar p$ are defined in the following way. We  first introduce the
1-form $\hat{p}=p_\mu dy^\mu$, and then we express it in terms of the
$z^a$ coordinates
\be
\hat{p}=\bar p_a dz^a =p_\alpha W^\alpha_a dz^a 
\ee
The prefactor of $\hat{C}^{(A)}(p)$ is fixed by our previous
normalization which coincides with the result for $p=0$.
Notice that the difference between $\f_A$ and $\f_B$ is order
$\epsilon$. Thus, if we neglect this term as higher order the result
simplifies and the real quadratic form in the exponent  becomes 
 \be
 \frac{\sqrt{N_1 N_2 \V}}{2\pi} \|p\|^2= \frac{\sqrt{N_1 N_2
 \V}}{2\pi} p_\alpha p_\beta \hat g^{\alpha \beta}
 \ee
\section{Computation to higher orders in $\epsilon$}
\label{higherorders}
The computation of the vector potential for the fractional instanton
can be continued to higher orders using essentially the same strategy
that was used for the calculation to order $\epsilon$. One has to
sequentially solve for $\W_{\mu , n}$ by the equivalent of the first
equation and then solve for $S^{(a)}_{\mu, n}$ by the equivalent of
the second equation. The procedure to solve for this second equation
would be based on the Fourier decomposition as before. This is an
inhomogeneous equation, where the known part involves the coefficients
determined already at lower orders. The treatment of the unknown term 
$\bar\eta^{b c}_d \partial_b S^{(a)}_{c, n}$ is done once more in a
quaternionic fashion, introducing the matrices $\sigma$ and $\bar\sigma$
and parameterizing $S^{(a)}_{c, n}$ as follows
\be
\label{eveneq}
\boldsymbol{\tilde S^{(a)}}_{, n} =\boldsymbol{\tilde{\partial}}
\boldsymbol{\tilde G}_{,n}
\ee
where now $\boldsymbol{\tilde G}_{,n}= G_{b,n} \sigma_b$. The equation then
involves the Laplacian of $G_{b,n}$, with Fourier coefficients 
easily expressible in terms of those of $G$. Again adding a constant 
term to $\tilde S^{(a)}_{c, n}$ gives also a solution of the equation,
however this is related space-time
translations (see our
discussion in section~\ref{deformation}). We fix the solution uniquely by setting this constant to
0. 

Now we consider the odd equations, which fix the $\W_{\mu, n}$. For
$n>0$ this now becomes inhomogeneous too. The part containing the
unknown to be determined looks like
\be
\bar\eta^{b c}_d \bar{D}_b \W_{c, n}
\ee
This must be expressed in quaternionic form 
\be
\boldsymbol{\bar{D}} \boldsymbol{\tilde{\W}}_{,n}
\ee
We now proceed to analyze the structure of the quaternionic operator
$\boldsymbol{\bar{D}}$.
Written in matrix form we have
\be
\begin{pmatrix}
\bar D_0+i \bar D_3 & i \bar D_1 + \bar D_2 \cr
i \bar D_1 -\bar D_2 & \bar D_0 -i \bar D_3
\end{pmatrix}
\ee
Notice that the commutator of the covariant derivatives is given in
terms of the abelian field. With our choices the 0 and 3 components
commute with the 1 and 2 components. Concerning the two combinations
one has 
\be
[\bar D_0+i \bar D_3, \bar D_0-i \bar D_3]=-\frac{4 \pi \f_A}{N}
\ee
This looks similar to the commutation relations of creation and
annihilation operators. Indeed, if we write 
\be
\mathbf{a}= i \sqrt{\frac{4 \pi \f_A}{N}} (\bar D_0+i \bar D_3)
\ee
the previous relation becomes exactly 
\be
[\mathbf{a}, \mathbf{a}^\dagger]=1
\ee
This formulation is particularly inspiring for physicists since we are
well acquainted with the properties of creation and annihilation
operators. In particular, the operator
$\mathbf{a}^\dagger \mathbf{a}$ has an  spectrum given by the positive
integers. The lowest eigenvalue is zero and its eigenvector is the
state annihilated by $\mathbf{a}$. A similar thing can be done for
$\bar D_1$ and $\bar D_2$ generating other creation-annihilation
operators $\mathbf{b}^\dagger$ and  $\mathbf{b}$, commuting with the
previous ones. If we write back the quaternionic operator in this
notation we have
\be
\boldsymbol{\bar{D}}= -i \sqrt{\frac{N}{4 \pi }} \begin{pmatrix}
 \mathbf{a}/\sqrt{\f_A} & i
\mathbf{b}^\dagger/ \sqrt{\f_B}\cr
i \mathbf{b}/\sqrt{\f_B}  &  \mathbf{a}^\dagger/\sqrt{\f_A}
\end{pmatrix}
\ee
It is now clear why, at the level of the first equation, the form 
of $\W_{\mu , 0}$ was taken in that particular fashion. The only
non-zero component was annihilated by $\mathbf{b} $ and  $\mathbf{a}$.
For $n>0$ however the creation operators also contribute. An
appropriate basis of the space is given by  the simultaneous eigenstates 
of the two number operators $\mathbf{a}^\dagger \mathbf{a}$ and 
$\mathbf{b}^\dagger \mathbf{b}$. Since these operators commute with
$\Op_\alpha$, it is enough to consider the $0-0$ element of the 
corresponding matrix and construct the complete $N_1\times N_2$ matrix
in the same way as we did for the calculation to order
$\sqrt{\epsilon}$. Thus, we define (up to a normalization)
\be
\mathbf{a}^\dagger \mathbf{a} \Psi(n,n',y) =n  \Psi(n,n',y)\  ; \quad
\mathbf{b}^\dagger \mathbf{b} \Psi(n,n',y) =n'  \Psi(n,n',y)
\ee
Then we can expand the solution $\bW_0$ to any order in $\epsilon$ as 
a linear combination 
\be
\bW_0(y)= \sum_{n,n'=0}^\infty c_{n n'}  \Psi(n,n',y)
\ee
It is not difficult to construct the basis functions $ \Psi(n,n',y)$ explicitly
starting from the solution for $n=n'=0$ that we used before. This can
be done with the well-known formulas for the harmonic oscillator 
\be
\Psi(n,n',y) =\frac{1}{\sqrt{n! n'!}} (\mathbf{a}^\dagger)^n
(\mathbf{b}^\dagger)^{n'} \Psi(0,0,y)
\ee
The procedure to  solve the corresponding  equation to higher
orders   is essentially the same detailed in the appendix of the
paper~\cite{GonzalezArroyo:2004xu}, but generalized from 2 dimensions to 4 dimensions.
For diagonal metric tensor the 4 dimensions split naturally into two
2-dimensional planes and the formulas can be obtained readily from that
reference. In the general case, the 4 dimensions are intermingled and
the formulas become more involved. The explicit formulas necessary to
implement the iterative procedure will be given
elsewhere~\cite{inpreparation}.   
\section{Conclusions}
\label{conclusions}
In this paper we have set up the formalism for writing analytic
formulas for the gauge potentials and field strength of (minimum action) fractional
instantons for SU(N) gauge theories on a 4 dimensional torus. We have
given the general solution of the  constant field strength type 
studied by `t Hooft~\cite{thooft}. We have clarified how self-duality 
implies conditions on the flat metric
tensor which determines the length and scalar product of the
generators of the lattice defining the torus, and expressed the
general solution for this metric. Other  metric tensors can be written
as deformations of these solutions. The self-duality condition then
gives rise to  a hierarchy of equations which allows to
determine the gauge potentials as a power series in the parameter
controlling the deformation. The study constitutes in itself a proof
that there are indeed non-constant fractional instantons for metrics
not too far from those giving directly self-duality. The whole
procedure generalizes the structure already devised for the SU(2)
case~\cite{su2paper}, and deals with the multiple complications
associated to the higher rank. 

The method allows multiple extensions of this work which have been
left out of this paper. First of all, one can set up a methodology to
extend the computation to higher orders. Something very similar was
already done in the case of two-dimensional abelian Higgs vortices in
Ref.~\cite{GonzalezArroyo:2004xu}. There we were able to go to  up to order $51$ in
the expansion. Here of course, everything becomes more complex, so
maybe one cannot go that high in the expansion. An alternative
possibility based on our construction is to use  a variational method.
The construction privileges a certain basis in the matrix functional
space. Then one could use a truncated basis space and determine the
optimal values of the parameters to minimize the anti-self-dual part
of the field strength. This method is worth being explored. 

The present work has concentrated on determining the self-dual
configuration  with minimal action in the sector with $Q=1/N$. The
reason is that this configuration is essentially unique up to
space-time translations and gauge transformations. Higher values of
the topological charge imply a much richer moduli space. These are
essentially multi-fractional instanton solutions. Their general
structure could be quite rich. Studying them with our method is
however feasible as was done in the simpler two-dimensional abelian
multi-vortex solutions~\cite{GonzalezArroyo:2004xu}. 

Having an analytic control of the  vector potential and field strength
associated to the fractional instanton opens the way to many
collateral analytic calculations. For example one can compute the
zero-modes of the Dirac equation both in the fundamental and in the
adjoint representations. In the former case, one can use these
solutions to construct the Nahm dual of the fractional instantons.
This might be particularly useful for the Nahm-self-dual cases. The
adjoint zero-modes might be useful in the context of Adjoint QCD with
its many attractive properties. Last but not least the formulas
developed here can be analysed to see simplifications occurring in
certain limits which might give rise to compact analytic expressions
and connections with other instanton solutions. Both the non-abelian
self-dual vortices~\cite{GonzalezArroyo:1998ez, Montero:1999by,
Montero:2000pb} and the
calorons~\cite{Lee:1998bb, Lee:1998vu, Kraan:1998kp, Kraan:1998pm} can be obtained as limiting cases of these
fractional instantons on the torus. The case of calorons is
particularly interesting as there are analytic solutions to which to
compare. 

\section{Acknowledgments}
I want to  thank Margarita Garc\'{i}a P\'{e}rez for useful discussions.
I  acknowledge financial support from the MINECO/FEDER grant
FPA2015-68541-P  and the Spanish Agencia Estatal de
Investigacion through the grant “IFT Centro de Excelencia Severo Ochoa
SEV-2016-0597”.
\bibliography{fractional}

\end{document}